\documentclass[usenatbib]{mnras}

\usepackage{newtxtext,newtxmath}
\usepackage[T1]{fontenc}
\usepackage{ae,aecompl}

\usepackage{graphicx}	% Including figure files
\usepackage{amsmath}	% Advanced maths commands
\usepackage{amstext}
\usepackage[usenames]{xcolor}
\usepackage{booktabs,gensymb}
\usepackage[referable]{threeparttablex}
\usepackage{hyperref}
\hypersetup{colorlinks=true, urlcolor=blue, citecolor=blue, linkcolor=blue}

\newcommand{\rev}[1]{#1}
\newcommand{\revv}[1]{#1}
\newcommand{\Msun}{\,M_{\odot}}
\newcommand{\Ms}{M_{*}}
\newcommand{\epsff}{\epsilon_{\mathrm{ff}}}

\title[Evolution of Disc Thickness]{Evolution of Disc Thickness in Simulated High-Redshift Galaxies}

\author[Meng et al.]{
Xi Meng\thanks{E-mail: xim@umich.edu}\href{https://orcid.org/0000-0002-8276-4164}{\includegraphics[scale=0.5]{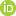}} and
Oleg Y. Gnedin \href{https://orcid.org/0000-0001-9852-9954}{\includegraphics[scale=0.5]{orcid.png}}
\\
Department of Astronomy, University of Michigan, Ann Arbor, MI 48109, USA
}

\date{Accepted XXX. Received YYY; in original form ZZZ}

\pubyear{2020}

\begin{document}
\label{firstpage}
\pagerange{\pageref{firstpage}--\pageref{lastpage}}
\maketitle

\begin{abstract}
 We study the growth of stellar discs of Milky Way-sized galaxies using a suite of cosmological simulations. 
 We calculate the half-mass axis lengths and axis ratios of stellar populations split by age in galaxies with stellar mass $\Ms=10^7-10^{10}\Msun$ at redshifts $z>1.5$. 
 We find that in our simulations stars always form in relatively thin discs, and at ages below 100~Myr are contained within half-mass height $z_{1/2}\sim$0.1~kpc and short-to-long axis ratio $z_{1/2}/x_{1/2}\sim$0.15. 
 Disc thickness increases with the age of stellar population, reaching median $z_{1/2}\sim$0.8~kpc and $z_{1/2}/x_{1/2}\sim$0.6 for stars older than 500~Myr.
 We trace the same group of stars over the simulation snapshots and show explicitly that their intrinsic shape grows more spheroidal over time.
 \revv{We identify a new mechanism that contributes to the observed disc thickness: rapid changes in the orientation of the galactic plane mix the configuration of young stars.
 The frequently mentioned "upside-down" formation scenario of galactic discs, which posits that young stars form in already thick discs at high redshift, may be missing this additional mechanism of quick disc inflation.
 The actual formation of stars within a fairly thin plane is consistent with the correspondingly flat configuration of dense molecular gas that fuels star formation.}
\end{abstract}

% Select between one and six entries from the list of approved keywords. Don't make up new ones.

\section{Introduction}

 The stellar disc of the Milky Way (MW) galaxy shows two distinct geometric components, the thin disc and the thick disc, with characteristic scale heights $\sim$0.3~kpc and 0.9~kpc, respectively \citep{Juric:2008aa}. 
 Similar thin disc plus thick disc structure is also observed in other nearby galaxies \citep[e.g.][]{Reddy:2006aa,Yoachim:2008aa,Yoachim:2008ab}. 
 The MW thin and thick discs differ in chemical abundances \citep{Bensby:2005aa} -- the thin disc being more metal rich with lower $[\alpha/{\rm Fe}]$, and the thick disc being more metal poor with higher $[\alpha/{\rm Fe}]$ -- suggesting that the thin disc is mainly composed of relatively young stars, while the thick disc mainly of old stars. 
 The typically quoted axis ratios of the thin and thick discs are $\sim0.1$ and 0.4, respectively \citep[e.g.,][]{Bovy:2013aa, Bensby:2011aa}.
 Results of massive spectroscopic surveys such as APOGEE suggested that instead of a clear separation between the thin and thick discs, the MW structure may be better described by a superposition of many mono-abundance populations, each with a single exponential scale height and scale length \citep{Bovy:2012aa, Bovy:2016aa}. 
 For older populations, indicated by lower metallicity and enhanced $[\alpha/{\rm Fe}]$, the scale heights monotonically increase with age (from $\approx$200~pc to 1~kpc), while the scale lengths decrease (from >4.5~kpc to 2~kpc). 

 The disc structure has also been studied at high redshift. Unlike the regular thin discs seen at low redshift, high-redshift galaxies show thick discs which are often clumpy \citep[e.g.][]{Elmegreen:2007aa,Overzier:2010aa,Swinbank:2010aa}. 
 Chain and spiral galaxies in the \textit{Hubble Space Telescope (HST)} Ultra Deep Field have the ratio of scale height to radial scale length $\sim 1/3$ \citep{Elmegreen:2006aa}.
 \citet{Elmegreen:2017aa} found the scale heights of galaxies in the \textit{HST} Frontier Fields Parallels to increase with galaxy mass and decrease with redshift.
 They noted that clump regions have smaller scale height, and that the thick disc is observed best between the clumps, with a larger scale height. 
 The overall evolution of galaxy shape appears to transform galaxies from prolate and spheroidal shapes at high redshift to thin discs at low redshift \citep{Law:2012aa,van-der-Wel:2014aa,Zhang:2019aa}. 

 Current cosmological simulations successfully reproduce the observations that the present day distribution of young stars is thinner than that of old stars, and that galaxy discs are thicker at high redshift \citep[e.g.][]{Bird:2013aa,Pillepich:2019aa,Buck:2020ab}. However, the origin of such a transition from thick to thin discs, and how stars form in thick discs at high redshift, is still actively debated.
 
 There are several scenarios that attempt to explain the formation of thick discs. 
 One posits that the old stellar populations formed thick and kinematically hot at high redshift \citep[e.g.][]{Tutukov:2000aa, Bird:2013aa, Stinson:2013aa}, as mono-age or mono-abundance populations in simulations show thicker and shorter discs at birth.
 Other scenarios suggest that stars always form in thin discs and are heated to larger heights because of scattering by massive clumps \citep{Bournaud:2009aa,Comeron:2011aa,Comeron:2014aa,Beraldo-e-Silva:2020aa} or radial migration \citep{Schonrich:2009aa, Loebman:2011aa}. 
 The third type of studies suggest that thick discs are related to galaxy mergers, including heating of a thin disc in a merger, direct accretion of stars from satellite galaxies, and star formation in merging gas-rich systems \citep[e.g.][]{Reddy:2006aa}. 
 Considering that mergers are more frequent at high redshift, these scenarios may lead to thicker discs of old stellar populations. 

 In this paper, we revisit the evolution of galactic disc thickness with stellar age, taking advantage of a recent suite of cosmological simulations of MW-sized galaxies with ultrahigh resolution.
 We test whether young stars actually form in thin discs and how quickly the scale height grows with stellar age. 
 In \autoref{sec:shape} we describe the simulations and how we calculate the shape of the simulated galaxies. 
 We present the evolution of thickness of our simulated galaxies with stellar age in \autoref{sec:result}.
 We compare our results with observations and other simulations and discuss the factors that contribute to the thickening of galactic discs in \autoref{sec:discuss}, and present our conclusions in \autoref{sec:conclude}.

\section{Shape of High-Redshift Galaxies} \label{sec:shape} 

\subsection{Simulation Suite}

 We use a suite of cosmological simulations run with the Adaptive Refinement Tree (ART) code \citep{Kravtsov:1997aa,Kravtsov:1999aa,Kravtsov:2003aa,Rudd:2008aa} and described in \citet{Li:2018aa} and \citet{Meng:2019aa}. 
 All runs start with the same initial conditions in a periodic box of 4 comoving Mpc, producing a main halo with total mass $M_{200}\sim10^{12}\Msun$ at $z=0$, similar to that of the MW. 
 The ART code uses adaptive mesh refinement to reach higher spatial resolution in dense regions. 
 The lowest resolution is set by the root grid, which has $128^3$ cells. 
 This sets the dark matter particle mass $m_{\rm DM}\approx 10^6\Msun$. 
 The finest refinement level is adjusted in runtime to keep the physical size of gas cells on that level between 3 and 6~pc. 
 Because of strong stellar feedback, few cells remain at this finest refinement level and the typical spatial resolution of molecular gas is 36-63~pc. 
 
 The simulations include three-dimensional radiative transfer using the Optically Thin Variable Eddington Tensor approximation \citep{Gnedin:2001aa} of ionizing and ultraviolet (UV) radiation from stars \citep{Gnedin:2014aa} and the extragalactic UV background \citep{Haardt:2001aa}, non-equilibrium chemical network that calculates the ionization states of hydrogen and helium, and phenomenological molecular hydrogen formation and destruction \citep{Gnedin:2011aa}. 
 The simulations incorporate a subgrid-scale model for unresolved gas turbulence \citep{Schmidt:2014aa,Semenov:2016aa}. 
 Star formation is implemented with the continuous cluster formation (CCF) algorithm \citep{Li:2017ab}, where each stellar particle represents a star cluster that forms at a local density peak and grows its mass via accretion of gas and star formation until feedback of its own stars terminates the growth. 
 The feedback recipe includes early radiative and stellar wind feedback, as well as a supernova (SN) remnant feedback model \citep{Martizzi:2015aa,Semenov:2016aa}. 
 The momentum feedback of the SN remnant model is boosted by a factor $f_{\rm boost}=5$ \rev{to compensate for the underestimation due to numerical discreteness of the isolated SN model} and to match the star formation history expected from abundance matching. 
 The simulations include several runs with different values of the local star formation efficiency (SFE) per free-fall time, which ranges from $\epsff=10\%$ to 200\%. 
 For a full description of the star formation and feedback recipe, see \citet{Li:2017ab,Li:2018aa} and \citet{li_gnedin19}. 

\rev{The simulations are run to $z\approx1.5$. The high-redshift galaxies in our simulations are turbulent and irregular. They show prolate shape and contain clumps of young stars, consistent with the observed high-redshift clumpy galaxies. We remind the reader that these galaxies are unlike the regular galaxies at low redshift with a thin rotating disc and central bulge. Instead, the high-redshift galaxies are rather more irregular and dominated by turbulence. 

Apart from the fiducial runs with $f_{\rm boost}=5$, our simulation suite also includes a weaker feedback run with $f_{\rm boost}=3$. The weaker feedback run produces a much higher stellar mass than expected from abundance matching and has a thin rotating disc even at high redshift. We do not consider that run as realistic and exclude it from analysis in this paper. More details about the structure of our simulated galaxies are given in \citet{Meng:2019aa} and \citet{Meng:2020aa}}

 In this paper we analyze four runs that used different values of $\epsff$. 
 The number after "SFE" in the run name corresponds to the percentage of local $\epsff$. 
 At redshifts $z\approx2-5$, there is one main galaxy and several smaller galaxies. 
 We investigate the structure of the main galaxy and other isolated galaxies with stellar mass above $10^7\Msun$. 
 
\subsection{Measurements of galaxy shape}

 To measure galaxy shape, first we need to define the extent of a central galaxy and eliminate possible effects of satellites and galaxy mergers. 
 For this purpose we calculate galaxy properties in a sphere centered on the galaxy center with radius 10~kpc for $z\leqslant5$ galaxies, and 5~kpc for $z>5$ galaxies. 
 This covers most of the stellar distribution of our galaxies. 
 
 We examined the robustness of this choice by evaluating several alternative radial cuts. 
 We determined the stellar extent of the galaxies by identifying a local minimum $r_{\rm M,min}$ in the spherically-averaged mass profile $dM(r)/dr$. If the profile decreases with $r$ and then begins to increase again, it means there is another stellar structure such as a satellite galaxy. 
 We use $r_{\rm M,min}$ as one of alternative radial cuts. 
 The median of these radii is about 10~kpc for $z\leqslant5$ galaxies, and 5~kpc for $z>5$ galaxies, which informs our final choice.
 The other alternative we tried is 20~kpc for $z\leqslant5$ galaxies, and 10~kpc for $z>5$ galaxies, which is close to the extent of the largest galaxies in high-redshift observations \citep[e.g.][]{Elmegreen:2017aa}.
 
 We found that for most of our simulated galaxies, choosing different maximum radius has little effect on the determination of galaxy orientation or stellar scale lengths along all axes.
 Using these alternative radial cuts results in the median fractional difference in the half-mass axis lengths smaller than 0.1~dex, and the orientation of the majority of galaxies changes by less than 1 degree.

 \begin{figure}
  \includegraphics[width=\columnwidth]{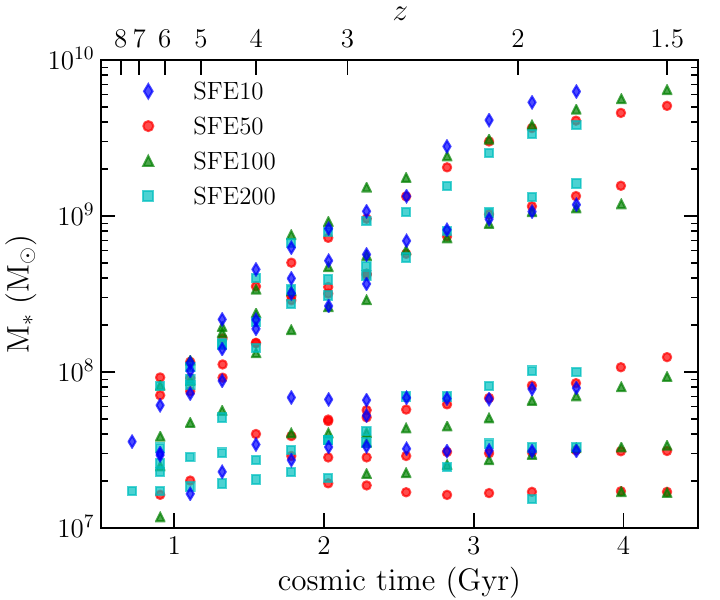}
  \vspace{-4mm}
  \caption{Stellar mass of simulated galaxies vs. cosmic time and redshift in the simulation snapshots selected for analysis. In addition to the main galaxy, each snapshot contains several smaller isolated galaxies.}
  \label{fig:overview}
 \end{figure}
 
 We study all isolated (also referred to as central to their dark matter halo) galaxies, which are located outside of virial radii of other galaxies. 
 The number of stellar particles in a galaxy ranges from a few hundred for galaxies with $\Ms\sim10^7\Msun$ to several million for galaxies with $\Ms\sim10^{10}\Msun$. We require galaxies to have $\Ms>10^7\Msun$ so that they contain \rev{at least 100 stellar particles for the shape analysis to be reliable.}
 
 For the purpose of comparison with observational samples, different stages in the evolution of a given simulated galaxy can be considered uncorrelated if they are separated by a long enough time.
 We choose the separation between simulation snapshots based on the dynamical time calculated at the typical largest extent of stellar distribution, which is $R/V_{\rm circ}\sim 60-100$~Myr.
 We wish to use the snapshots separated by at least two dynamical times.
 The time interval between our snapshots is $\sim$150~Myr at $z<5$ and $\sim$100~Myr at higher redshift, thus we select one in every two snapshots. 
 This results in separation of $200-300$~Myr between the analyzed epochs.

 We show the galaxy stellar mass and cosmic time of our sample in \autoref{fig:overview}. 
 Each point corresponds to an isolated galaxy at a selected snapshot. 
 One can see two relatively large galaxies and several smaller galaxies at these epochs.
 
  \begin{figure}
  \centering
  \includegraphics[width=0.9\columnwidth]{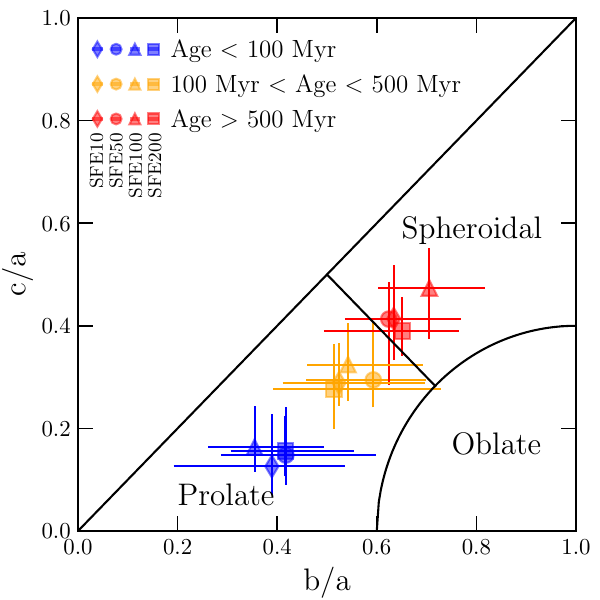}
  \vspace{-1mm}
  \caption{Intrinsic axis ratios of stellar populations of different age. Points with error bars show the median and interquartile ranges of stars in a given age bin for all galaxies in our sample. For most galaxies, the stellar distribution transitions from prolate to more spheroidal shape as stellar age increases.}
  \label{fig:sum_ageaxisratio}
 \end{figure}
 
 High-redshift galaxies often show irregular shapes unlike axisymmetric discs observed at low redshift, and they are often dominated by turbulent motions instead of rotation. Therefore, following \citet{Meng:2019aa} we use the shape tensor \citep{zemp_etal11} to determine the orientation of our simulated galaxies: 
 $$ \textbf{I} \equiv \sum_{k,i,j} M_k \, r_{k,i} \, r_{k,j}\ \textbf{e}_{i} \otimes \textbf{e}_{j} ,$$ 
 where $M_k$ is the mass of $k$-th stellar particle, $r_{k,i}$ are its coordinates in the galactocentric reference frame $(i=1,2,3)$, and $\textbf{e}_i$ are the three unit vectors of the coordinate axes. 
 The tensor can be diagonalized by a rotation matrix to obtain the principal moments of inertia $I_1 \geqslant I_2 \geqslant I_3$. 
 From these we calculate the axis ratios $b/a \equiv \sqrt{I_2/I_1}$ and $c/a \equiv \sqrt{I_3/I_1}$, which describe the shape of the mass distribution. 
 The orientation of the galaxy plane is given by the eigenvector corresponding to the smallest eigenvalue.
 Following \citet{van-der-Wel:2014aa}, we divide the parameter space of the axis ratios $b/a$ and $c/a$ into three distinct shapes of an ellipsoid: oblate, spheroidal, and prolate (elongated).

 \begin{figure}
  \includegraphics[width=\columnwidth]{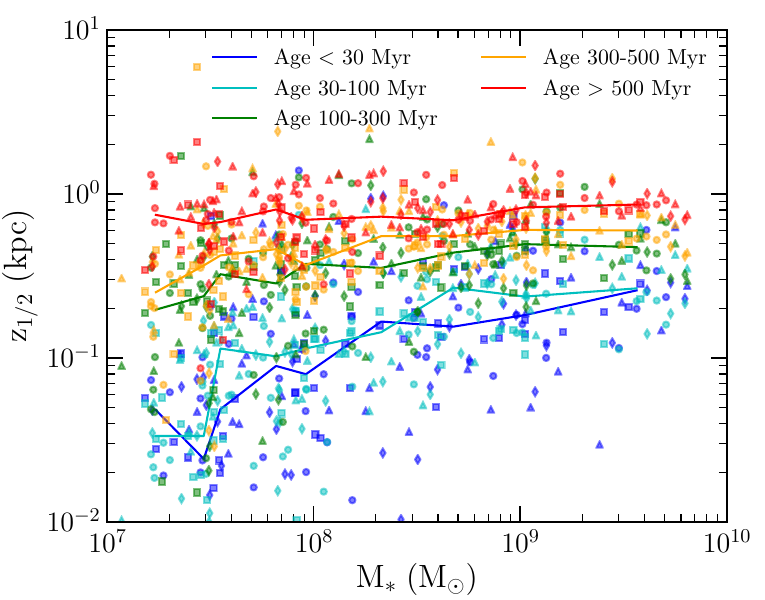}
  \vspace{-4mm}
  \caption{Disc half-mass height of stellar populations split by age. Galaxies in the four runs are shown by the same symbols as in \autoref{fig:overview}. Lines show median values of $z_{1/2}$ in bins of galaxy mass. For the youngest stars the disc height increases with galaxy mass, but at ages above 100 Myr the median height is independent of mass.}
  \label{fig:sumzhalf_Mstar}
 \end{figure}
 
 In \autoref{fig:sum_ageaxisratio} we show the evolution of the axis ratios with stellar age. 
 We separate stellar populations in three groups: less than 100~Myr, between 100 and 500~Myr, and older than 500~Myr. 
 Each point shows the median axis ratios of stars in an age group for all galaxies in our sample, and the errorbars show the interquartile ranges of the $c/a$ and $b/a$ distributions. 
 In most galaxies, $c/a$ and $b/a$ both increase with age, and the morphology of the stellar populations transitions from prolate to more spheroidal or triaxial shape. 
 Note that for this plot we calculated the shape tensor of each age group separately, so that the intrinsic orientation is possibly different for each group. Our goal here was to show that older stars have more spheroidal distribution even when the coordinate axes are chosen to minimize the short axis ratio $c/a$.

 For all subsequent analysis we fix a single orientation for a given galaxy, as it would be assigned in observation. We choose the plane orientation based on the shape tensor of stars younger than 100~Myr, because they contribute most of the observable rest-frame UV light. 
 We then transform the coordinates of stellar particles into the new coordinates given by the adopted orientation. 
 
 We use half-mass axis lengths to describe the thickness of different components of our galaxies. 
 The half-mass axis length is calculated as the length that contains half the cumulative mass along a given direction: $M(|x|<x_{1/2}) \equiv \frac{1}{2}M_{\rm tot}$, where $x$, $y$ and $z$ correspond to the long, middle, and short axes. 
 We use $z_{1/2}/x_{1/2}$ as a proxy for the ratio of scale height to scale length, and refer to it as "disc thickness".
 \rev{Note here we use "disc thickness" only to describe the thickening of stellar distributions as in observations. We do not distinguish between a flattened stellar distribution or a rotationally-supported disc. }
 
 In previous papers \citep{Meng:2019aa,Meng:2020aa} where we studied the relation between the star formation rate and interstellar medium, we defined galaxy plane as the orientation of the short axis of the shape tensor of neutral gas. Here we use the definition based on stars to match the observations of high-redshift stellar populations.
 We checked the difference in these two definitions of galaxy orientation: the median difference is about 30\degree\ for all galaxies in our sample. For the most massive galaxies, the difference is smaller, with a median of 16\degree. 

 \begin{figure}
  \includegraphics[width=\columnwidth]{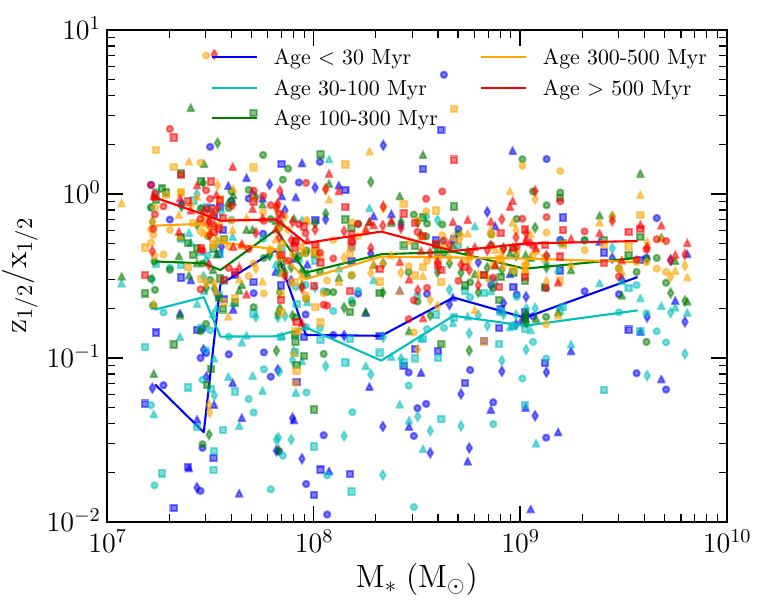}
  \vspace{-4mm}
  \caption{Ratio of the short to long axes, split by stellar age. Symbols are as in \autoref{fig:sumzhalf_Mstar}. The median disc thickness shows little variation with galaxy mass.}
  \label{fig:sumz2x_Mstar}
 \end{figure}
 
\section{Evolution of disc thickness} \label{sec:result}

 For a more detailed analysis of the thickness of stellar populations, we group stars into more fine age bins: <30~Myr, 30--100~Myr, 100--300~Myr, 300--500~Myr, >500~Myr. We then calculate the half-mass length for each age bin in the three established coordinate directions. 

 First we explore whether the thickness of stellar discs depends on galaxy mass, which itself increases over time. \autoref{fig:sumzhalf_Mstar} shows the half-mass height $z_{1/2}$ for stellar populations of different age. Each galaxy is represented by five small points corresponding to the heights of its five age populations. The shape of the symbols indicates from which run they were selected, but there is no systematic dependence in any of the results on the SFE adopted in the simulations. 
 
 Big diamonds show median values of the small points in galaxy mass bins, to allow better inspection of any systematic trends. The clearest trend is the half-mass height monotonically increasing with age of the stellar population. We describe it in more detail in the following plots. Any dependence on galaxy stellar mass is much less obvious. There may be a slight trend for the height of the youngest stars to increase with mass, but not for stars older than 100~Myr. Considering that the low-mass galaxies contain fewer than $10^3$ stellar particles, and even less in a given age bin, this trend may not be significant.
 
 \autoref{fig:sumz2x_Mstar} shows disc thickness as the ratio of the short to long axis lengths $z_{1/2}/x_{1/2}$. Similar to the half-mass height, the thickness also increases with the age of the stellar populations. The only possible exception is with the two youngest age bins, which show more scatter and variation at the low mass end. This is because the orientation we choose to define the disc plane is based on the shape of stars younger than 100~Myr, which includes both age bins. This definition automatically minimizes the thickness of both stellar populations. Depending on stellar mass and specific star formation history, either the $<30$~Myr bin or the 30-100~Myr bin may display a thinner disc. Therefore, we find little difference in all of our results between these two bins.
 
 A small fraction of points lie above the line $z_{1/2} = x_{1/2}$. As our disc orientation minimizes the thickness of stars younger than 100~Myr, some older stellar populations that formed in a very different orientation may show $z_{1/2} > x_{1/2}$. This includes even one of the younger age bins (<30~Myr or 30--100~Myr) for a few galaxies. Our definition of the disc orientation guarantees that at least one of the younger bins will have $z_{1/2}/x_{1/2}<1$, and for majority of the galaxies both bins do. However, a few cases with rapidly precessing disc plane show the other young bin with $z_{1/2} > x_{1/2}$. The rapid precession of the plane of young stars is illustrated later in \autoref{fig:anglechange}. Such precession may be caused by the clumping of star forming regions and the cycles of gas infall and expansion due to strong stellar feedback.
 
 For the oldest bin with age >500~Myr the ratio $z_{1/2}/x_{1/2}$ decreases slightly with increasing stellar mass, from about 0.75 at $M_*\sim10^7\Msun$ to about 0.5 at $M_* > 10^9\Msun$. \rev{A linear fit between log($z_{1/2}/x_{1/2}$) and log$\Ms$ has a slope -0.102$\pm$0.023.} Thus disc thickening appears to be more pronounced for the older stars in dwarf galaxies. Still the trend is weak and does not appear to be related to the velocity dispersion of gas from which the stars form. We find the median ratio of the velocity dispersion to galaxy circular velocity is roughly constant $\sigma_{g}/v_{\rm circ}\approx 0.6$ at all galaxy masses. The more likely cause of higher $z_{1/2}/x_{1/2}$ in dwarf galaxies is just the small number of young stellar particles, which leads to a more stochastic galaxy orientation.
 
\begin{figure}
  \includegraphics[width=\columnwidth]{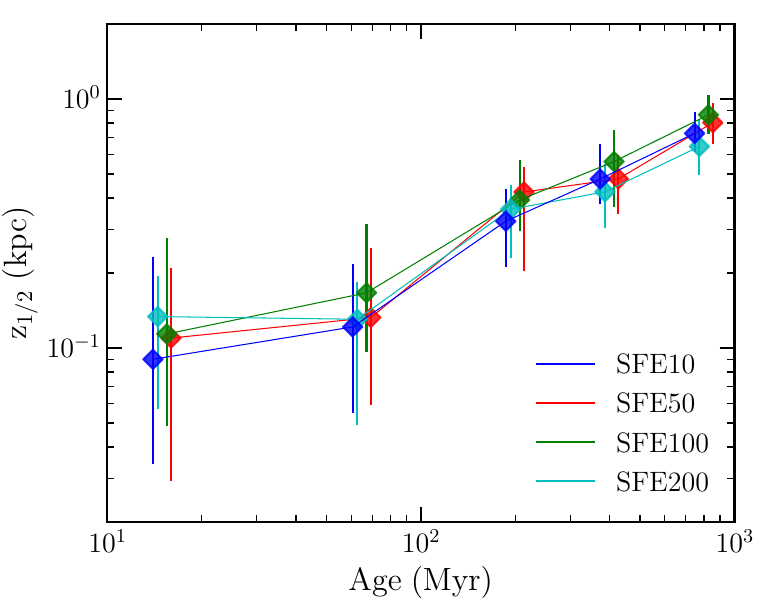}
  \vspace{-4mm}
  \caption{Evolution of the disc half-mass height with stellar age. Median $z_{1/2}$ of stellar populations in a given age bin for all galaxies are shown as big diamonds and connected with solid lines for clarity, analogously to \autoref{fig:sumzhalf_Mstar}. Instead of individual points, vertical errorbars show the interquartile range. The value of age is taken to be the median in a bin: about 15, 65, 200, 400, 800~Myr, respectively. A small offset in age is added to distinguish multiple points. Disc scale height monotonically increases with stellar age. There is no clear trend with the value of SFE used in the simulations.}
  \label{fig:sum_zhalf}
 \end{figure}

 Both the disc height and thickness increase with stellar age. \autoref{fig:sum_zhalf} illustrates this explicitly for the half-mass height. Instead of individual points for each galaxy's populations, vertical errorbars indicate the interquartile range of the galaxy distribution. Diamonds show the median values for all galaxies in our sample in a given age bin. We can see that the disc thickness clearly increases from $\sim$0.1~kpc at formation to $\sim$0.8~kpc after 1~Gyr. The scatter in the galaxy sample is approximately constant at 0.15-0.2~kpc, although it appears larger for younger ages due to the logarithmic scale.
 
\begin{figure}
  \includegraphics[width=\columnwidth]{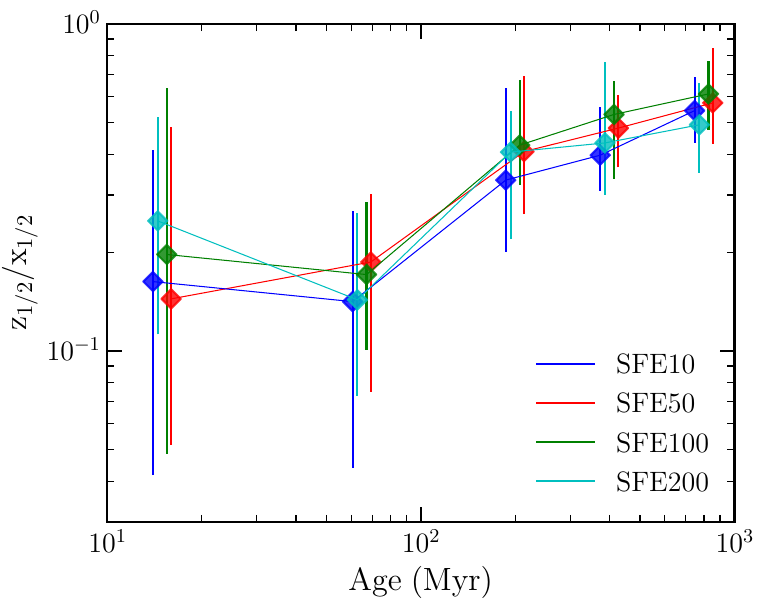}
  \vspace{-4mm}
  \caption{Evolution of the half-mass short-to-long axis ratio $z_{1/2}/x_{1/2}$ with stellar age. Symbols are as in \autoref{fig:sum_zhalf}. The median disc thickness clearly increases with age.}
  \label{fig:sum_z2x}
 \end{figure}

 Similarly, \autoref{fig:sum_z2x} shows that the half-mass ratio $z_{1/2}/x_{1/2}$ steadily increases with stellar age. Young stars always form in relatively thin discs with $z_{1/2}/x_{1/2}\lesssim0.2$, while after several hundred Myr the stars occupy significantly thicker shapes with $z_{1/2}/x_{1/2}\approx0.6$. Some runs have $z_{1/2}/x_{1/2}$ larger in the first age bin than in the second, again because the galaxy orientation is based on all stars younger than 100~Myr. Overall, both young stellar populations appear to occupy a similarly thin disc. As in the previous plots, there is no clear difference for the simulations with different adopted SFE.
 
 We provide the medians and interquartile ranges of $z_{1/2}/x_{1/2}$ for stellar populations of different age in \autoref{tab:z2x_iqr}.
 
 \begin{table*}
  \centering
  \begin{tabular}{lccccc}
   \toprule 
    \multicolumn{6}{c}{25--50--75\% range of $z_{1/2}/x_{1/2}$} \\
    Run & Age < 30~Myr & Age 30--100~Myr & Age 100--300~Myr & Age 300--500~Myr & Age > 500~Myr \\
   \midrule 
   SFE10  & 0.04 -- 0.16 -- 0.41 & 0.04 -- 0.14 -- 0.27 & 0.20 -- 0.33 -- 0.64 & 0.31 -- 0.40 -- 0.56 & 0.43 -- 0.54 -- 0.69\\
   SFE50  & 0.05 -- 0.14 -- 0.48 & 0.08 -- 0.19 -- 0.30 & 0.26 -- 0.41 -- 0.69 & 0.36 -- 0.48 -- 0.61 & 0.43 -- 0.57 -- 0.84\\
   SFE100 & 0.05 -- 0.20 -- 0.64 & 0.10 -- 0.17 -- 0.29 & 0.32 -- 0.43 -- 0.68 & 0.34 -- 0.53 -- 0.67 & 0.48 -- 0.61 -- 0.77\\
   SFE200 & 0.11 -- 0.25 -- 0.52 & 0.07 -- 0.14 -- 0.26 & 0.22 -- 0.41 -- 0.54 & 0.30 -- 0.43 -- 0.77 & 0.35 -- 0.49 -- 0.66\\
   \bottomrule 
  \end{tabular}
  \caption{Medians and interquartile ranges of the half-mass short-to-long axis ratios of stellar populations in age bins.} \label{tab:z2x_iqr} 
 \end{table*}
 
 To study the evolution of thickness of individual stellar populations, we trace the shape of the same group of stars through time. We choose all stars younger than 100~Myr within 10~kpc from galaxy center, for galaxies in the redshift range $z=2-4.6$. We follow these same stars in all consecutive snapshots and calculate the intrinsic half-mass axis ratio, in the coordinate frame defined by their own shape tensor. Note that this is different from the previous plots where the orientation of all stars is defined by the shape tensor of stars younger than 100~Myr at that snapshot. We choose the intrinsic orientation for this calculation because the galaxy orientation changes over time, and simple misalignment with current plane may lead to thicker inferred shape of the tracked stars. We stop tracking the stars when their host galaxies merge into larger halos, since mergers significantly alter the original shape of the stellar distribution.
 
 \autoref{fig:sumtrace} shows the resulting distribution for all selected populations. In total we have stellar groups selected from 118 independent galaxy snapshots. We only include groups with at least 100 stellar particles to be able to calculate the shape reliably. The stellar populations start from median age of $\sim$50~Myr and are traced in every consecutive snapshot after selection. The median value of $z_{1/2}/x_{1/2}$ increases from 0.15 when the stars are youngest to about 0.5 after 2~Gyr. The thickening of their distribution is unambiguous. This process saturates after 1-2~Gyr and therefore it is easy to miss in studies that have coarser time resolution. We do not find any dependence in this evolution on galaxy mass.
 \rev{Note that in \autoref{fig:sumtrace} the ratio $z_{1/2} / x_{1/2}$ is measured in the coordinate system defined by the shape tensor of the same stars that we trace. If instead $z_{1/2} / x_{1/2}$ is  measured in the orientation of the galaxy defined by young stars in each snapshot, the median of $z_{1/2} / x_{1/2}$ will increase to 0.5 in 1~Gyr, and further to 0.58 in 2~Gyr. This is because different stellar populations do not always align. We describe the orientation of different stellar populations in more detail in \autoref{fig:anglechange}.}

\section{Discussion} \label{sec:discuss} 

\subsection{Comparison with observations of high-redshift galaxies}

 When galaxies are observed edge-on, their vertical heights can be directly measured. 
 At $z>1.5$, optical and near infrared wavebands correspond to the rest-frame UV, which is contributed mostly by young stars. 
 \citet{Elmegreen:2006aa} fitted the vertical profiles of galaxies in \textit{HST} UDF with the functional form
 sech$^2(z/z_0)=4/[{\rm exp}(z/z_0)+{\rm exp}(-z/z_0)]^2$, where the half-light height $z_{1/2}$ is related to the scale height $z_0$ by $z_{1/2}=0.55\,z_0$. 
 Their results show that the half-light height of clumpy galaxies in the F850LP band to be $z_{1/2}=0.5\pm$0.2~kpc.
 This is larger than the thickness of the youngest stellar population of our galaxies: $0.1-0.2$~kpc. 
 Similarly, \citet{Elmegreen:2017aa} measured the sech$^2$ scale height of the high-redshift galaxy discs in the \textit{HST} Frontier Fields Parallels.
 They divided the galaxies into spiral type and clumpy (or "chain") type, and found that spiral galaxies are only present at $z<1.5$. 
 Since our simulated galaxies are at $z\geq1.5$ and appear clumpy, we compare our results to the clumpy/chain galaxies.
 For these systems \citet{Elmegreen:2017aa} find that the height is smaller for the star-forming clumps and larger in the interclump regions.
 They interpret it as the clumps representing the thin disc, while the interclump regions being more similar to the thick disc. 
 Therefore, the average scale height quantifies the thin disc and the maximum scale height in each galaxy quantifies the thick disc. 
 The median of the average half-light height of the clumpy galaxies in the redshift range 1.5-2.5 is $z_{1/2}=0.35\pm0.13$~kpc.
 This is closer to, but still larger than, the median half-mass height of the youngest stars in our simulated galaxies (see \autoref{fig:sumzhalf_Mstar}).
 Note that the observational sample includes more massive galaxies than in our simulations, from $10^9\Msun$ up to $10^{10.7}\Msun$, whereas our sample is below $10^{9.8}\Msun$. In the \citet{Elmegreen:2017aa} sample the disc height strongly increases with galaxy mass, which may partly account for the larger values in the observational sample.
 In contrast, the median of the maximum heights is even higher at $0.58\pm0.24$~kpc, well above our results.

 \begin{figure}
  \includegraphics[width=\columnwidth]{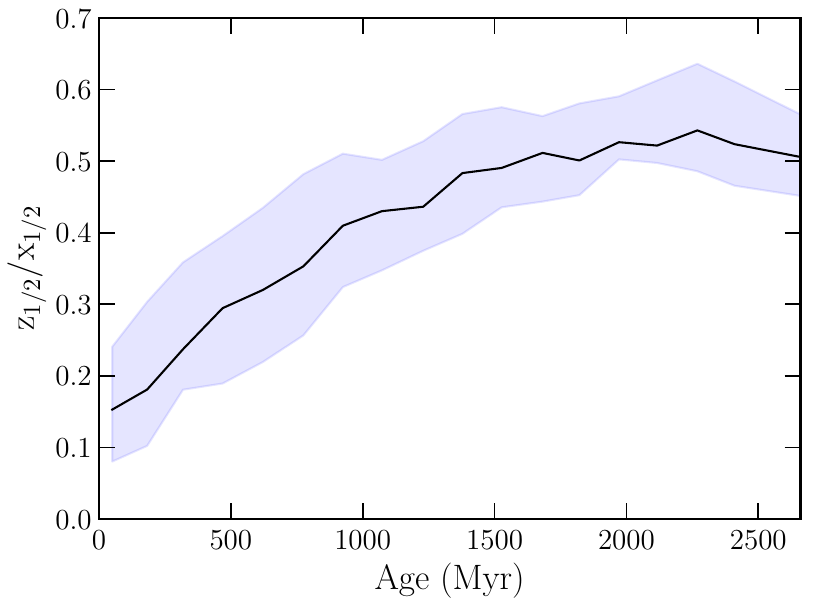}
  \vspace{-4mm}
  \caption{Evolution of the intrinsic short-to-long axis ratio of the same group of stars. Each group is selected as stars younger than 100~Myr in independent snapshots and traced over all subsequent snapshots until the galaxy mergers with a larger galaxy or reaches the end of the simulation. The line shows the median value for all combined groups in age bins, while the shaded region shows the interquartile range.}
  \label{fig:sumtrace}
 \end{figure}

 Although stars form in relative thin configuration in our simulated high-redshift galaxies, this configuration is still somewhat thicker than local star-forming discs, which typically have axis ratios below 0.15. 
 As a comparison, in our simulations the median short-to-long axis ratio of young stars is 0.15-0.2.
 High-redshift observations directly support thicker star-forming discs.
 For example, \citet{Reshetnikov:2003aa} find that in the Hubble Deep Field galaxies the ratio of disc scale height to scale length increases by a factor of at least 1.5 from low redshift to $z>1$. 
 \citet{Elmegreen:2006aa} quote the mean ratio of the radial half-length between clumps to sech$^2$ scale height to be 3.4$\pm$1.6, which corresponds to $z_{1/2}/x_{1/2}\gtrsim 0.16$, although exact conversion is not possible.
 This ratio is similar to that for young stars in our simulated galaxies.

 Besides the direct way of measuring the thickness of edge-on discs, there are also studies of shapes of high-redshift galaxies with other orientations.
 If galaxies are randomly projected on the sky, the observed shape distribution can be modeled to infer the intrinsic 3D shape distribution.
 \citet{Elmegreen:2005aa} found the intrinsic thickness of spiral galaxy discs in the \textit{HST} UDF to be $\approx$0.2-0.3. 
 Other studies divide high-redshift galaxies into mass and redshift bins and reconstruct the shape distribution for each bin to analyse the evolution of intrinsic galaxy shapes. 
 These studies found that high-redshift galaxies are more prolate or spheroidal than oblate \citep[e.g.][]{Law:2012aa}, and that the prolate fraction decreases with decreasing redshift and increasing galaxy mass \citep[e.g.][]{van-der-Wel:2014aa,Zhang:2019aa}. 
 For the mass and redshift range that best matches our simulated galaxies, $M_*=10^9-10^{11}\Msun$ and $z=1.5-3.6$, \citet{Law:2012aa} found galaxy minor/major and intermediate/major axis ratios $c/a\sim0.3$ and $b/a\sim0.7$. 
 \citet{Zhang:2019aa} give the average ratios $c/a\sim0.3$ and $b/a\sim0.5$ for galaxies with mass $10^{9.0-9.5}\Msun$ at $1.5<z<2.5$.
 This puts the intrinsic shape of observed high-redshift galaxies in the prolate regime, similar to our simulated galaxies.
 
 \begin{figure}
  \includegraphics[width=\columnwidth]{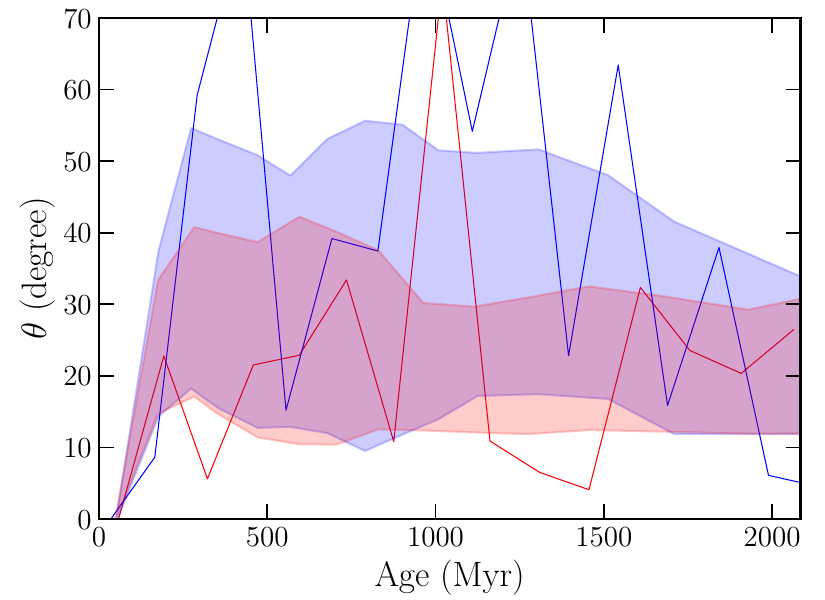}
  \vspace{-4mm}
  \caption{Angle between the orientation (normal) of young stars in a galaxy and the orientation of the group of stars chosen from this galaxy for \autoref{fig:sumtrace}, as a function of the age of the tracked stars. This angle is between $0\degree$ and $90\degree$ and always starts from $0\degree$ in the snapshot when the chosen group itself defines the orientation of the galaxy. The red/blue shaded regions show the interquartile range for galaxies with stellar mass larger/smaller than $10^8\Msun$ (roughly dividing the main progenitor branch and future satellites -- see \autoref{fig:overview}). Thin red and blue lines trace individual example galaxies in each mass bin. Age bins are selected to contain equal number of points, and are different from \autoref{fig:sumtrace} because splitting by mass reduces the number of points.}
  \label{fig:anglechange}
 \end{figure}
 
 \citet{Elmegreen:2017aa} also find the scale height of the most massive ($\Ms = 10^{9.5}-10^{10.5}\Msun$) clumpy galaxies to increase with galaxy mass and decreasing redshift. This trend is present for the youngest stars in our galaxies (\autoref{fig:sumzhalf_Mstar}).

\subsection{Comparison with simulations}
 
 Numerical simulations have also investigated the evolution of disc thickness. 
 For example, the Illustris TNG50 simulation suite has a much larger number of galaxies but lower spatial resolution (100-300~pc) than our simulations.
 In TNG50 the half-light height of star-forming galaxies with $\Ms=10^7-10^{10}\Msun$ in the rest-frame V band at $z=2$ is 200-400~pc \citep{Pillepich:2019aa}.
 Since the V band light is produced by stars of a wide range of age, this result should be compared to our age bins up to 500~Myr. Indeed, it agrees with the height of 100-500 Myr old stars in our galaxies.
 \citet{Pillepich:2019aa} define disc thickness somewhat differently from us, as the ratio of the half-mass height to the 2D circularized half-mass radius, $z_{1/2}/r_{1/2}$, but the difference should be limited to tens of percent.
 For galaxies in the mass range $10^8-10^9\Msun$ at $z=1.5-2$, the ratio $z_{1/2}/r_{1/2}$ ranges from 0.15 to 0.35. More massive galaxies, with $10^9-10^{10}\Msun$, are significantly less thick: $z_{1/2}/r_{1/2} \approx 0.1-0.2$.
 While these values are comparable to the average thickness of the 100-300~Myr old populations in our galaxies, we do not find as strong a dependence on galaxy mass. This difference may provide a useful observational test of the models.
 
 The TNG50 results were not split by stellar age, but they provide another measure of height for the H$\alpha$ emitting gas, which is a proxy for star formation rate and may indicate future location of very young stars. In all the TNG50 galaxies the height of the H$\alpha$ gas is consistently smaller than that of the V-band stars, which indicates the age trend that we described above for our results.
 
 In addition to measuring the disc height and length, \citet{Pillepich:2019aa} studied the intrinsic shape of  the stellar distribution in elliptical shells.
 They find that elongated (or prolate) galaxies are more common at higher redshift and lower galaxy mass: at $z>2$ about half of the galaxies with $\Ms=10^7-10^9\Msun$ have prolate shapes while the other half show spheroidal shapes.
 It is similar to our results (\autoref{fig:sum_ageaxisratio}) for the >100~Myr old stars.
 None of the TNG50 stellar distributions are as prolate as our young stars (<100~Myr old). The only comparably small values of $b/a$ are shown by the distribution of H$\alpha$ emitting gas, which may be a proxy for the very young stars. Therefore it is possible that in the TNG50 simulation stars also form only in thin discs. A more direct investigation of this trend would be desirable.
 
 In other suite of Vela simulations, \citet{Ceverino:2015aa} investigated the shape of high-redshift galaxies by fitting an ellipsoid to the 3D isodensity surface around the half-mass radius.
 They show one example of a galaxy at $z$=2.2 with $\Ms \approx 10^9\Msun$ that has the axis lengths of the stellar distribution equal to $(a, b, c)=(2.2, 0.76, 0.67)$~kpc. 
 Both the short axis length and the short-to-long axis ratio are similar to our results (\autoref{fig:sum_ageaxisratio}).

 A number of simulation studies favor the "upside-down" formation scenario for galactic discs, wherein stars form in thick discs at high redshift with a kinematically "hot" configuration, and then in progressively thinner discs at lower redshift. 
 For example, \citet{Buck:2020ab} used the NIHAO-UHD simulations of isolated MW-sized galaxies to investigate the evolution of height of stellar populations split in age bins of $\Delta t=$2~Gyr.
 They find that the exponential scale heights at birth decrease with stellar age, i.e. the "upside-down" formation.
 They also find that the scale heights at present day are larger than those at birth. This suggests that the disc thickness today is set partly at the time of formation and partly by subsequent secular heating.
 \revv{Note that unlike our globally measured $z_{1/2}$, \citet{Buck:2020ab} excluded the central galaxy and calculated the exponential scale height in the radial range $2<R<25$~kpc. The numbers they find for $\sim$10~Gyr old stars (i.e. formation at $z\approx 1-2$) are $z_d=0.8-1.5$~kpc, which correspond to the half-mass height $z_{1/2}=z_d\ln2 \approx 0.5-1.0$~kpc.}
 This is within the range of thickness of the oldest stars in our galaxies. 
 However, these estimates are likely to be incorrectly classified as "at birth". \citet{Buck:2020ab} choose wide age bins of $\Delta t=$2~Gyr, during which the disc orientation can evolve significantly as shown by our results, and therefore the stellar thickness may already be artificially inflated. This is supported by inspecting their Figs. 13 and 14, which show 3 of 5 galaxies with larger scale height than scale length "at birth" for stars older than 10 Gyr. This confirms the disc precession which hides the actual thickness of the star-forming disc. A finer time resolution is required to uncover the trend indicated by our results.

 Other simulation studies that favor the "upside-down" disc formation scenario also take wide 1-2~Gyr age bins for mono-age stellar populations \citep[e.g.][]{Bird:2013aa, Stinson:2013aa, Ma:2017aa, Navarro:2018aa}. They consistently find that old stars form in structures that are radially compact and relatively thick, while younger stars form in progressively larger, thinner, and colder configurations.
 \revv{We show that this cannot be used as confirmation of the "upside-down" scenario.}
 The stars could either form in actually thicker discs or form in thin discs and get quickly displaced from the changing midplane. This is especially important at high redshift where more frequent mergers may change the galaxy orientation more often. Only tracing the stellar population from the very young stars can distinguish between these two cases.
 \revv{For example, a recent paper by \citet{Bird:2020aa} analysed the upside-down formation scenario in simulated galaxies with smaller time intervals (50~Myr), illustrating both effects of the disc forming kinematically hotter at high redshift, and the stars being heated after birth. Since their analysis covers later epochs ($z<1.5$) than our simulations and quantifies only the velocity dispersion instead of disc thickness, we cannot make a direct comparison with their results.}

\subsection{On the cause of thickening of galactic discs}

 There are many physical mechanisms that lead to real thickening of galactic discs with fixed orientation. \rev{Classic papers by \citet{Spitzer:1951aa} and \citet{Spitzer:1953aa} pointed out that gravitational scattering by massive gas clouds could increase the velocity dispersion of stars in the galactic plane, especially considering differential galactic rotation. \citet{Lacey:1984aa} generalized the calculation by considering the vertical motion of stars and found that the velocity dispersion increases steadily while the shape of the velocity ellipsoid remains fixed. More recently,} \citet{Bournaud:2009aa} found that secular internal processes, such as gravitational instabilities and scattering by large clumps, lead to increased but constant disc height over galactocentric radius, while galaxy mergers lead to flaring of discs at large radii. 
 \citet{Grand:2016aa} showed the galactic bar and tidal perturbations from satellites to be responsible for most of the disc heating at $z\lesssim1$ in the Auriga simulations. 

 In addition to these real processes, we have identified another important, and possibly dominant, mechanism leading to apparently larger axis ratios of high-redshift galaxies. It is the rapidly changing orientation of the disc midplane.
 For example, the intrinsic axis ratio for the same stellar population 1~Gyr since birth is only $\sim$0.4 (\autoref{fig:sumtrace}), while the observable axis ratio for the oldest stars with the median age 800~Myr is already $\sim$0.6 (\autoref{fig:sum_z2x}).
 The difference comes from the different definitions of the galaxy orientation. In observations, only one orientation exists in the plane of the sky, and because of redshifting of starlight, it is defined by the distribution of blue young stars. Because of rapid midplane precession, this orientation may not align with the intrinsic orientation of older populations and artificially inflate the measured thickness.
 
 \autoref{fig:anglechange} illustrates the precession of the disc plane directly. The angle between the short axis of the configuration of tracked stars for \autoref{fig:sumtrace} and the short axis defined by the current young stars rises up to $\sim$20\degree-40\degree\ in less than 200~Myr and then oscillates around those values. The path of individual galaxies has a much wider variation than the interquartile trend, as illustrated by two cases in the figure.
 
 The median angle offset is noticeably larger for low-mass galaxies, $\Ms < 10^8\Msun$. Less massive galaxies are more easily affected by accretion and mergers which can change the disc orientation. Also, the determination of the plane is more stochastic for low-mass galaxies simply because of the smaller number of stellar particles.
 
 Overall, the plane precession is a significant contributor to the thicker appearance of high-redshift discs. Investigation of such offset at lower redshifts will be important for disentangling this effect from the real disc expansion by secular and external heating. Unfortunately, our current simulation suite does not extend beyond $z\approx 1.5$. We plan to investigate this issue with new simulations in progress and encourage other groups with available simulations to look into it as well.

\section{Conclusions} \label{sec:conclude} 

 We investigated the thickening of high-redshift galaxy discs using a suite of ultrahigh-resolution cosmological simulations. We selected a galaxy sample with stellar mass $\Ms = 10^7-10^{10}\Msun$ at redshifts $z\geq1.5$. We calculated the half-mass disc height and the axis ratio for stellar populations split into narrow age bins, from $<30$~Myr to $>500$~Myr. The main results are summarized below: 
 \begin{itemize}
  \item Using the shape tensor, we find the intrinsic 3D shape of high-redshift galaxies to be prolate or elongated, unlike the local axisymmetric discs. Stars younger than 100~Myr are confined to a more prolate and thinner configuration, while older stars gradually transition to a spheroidal shape (\autoref{fig:sum_ageaxisratio}).

  \item Young stars in our simulated galaxies always form in thin discs, with half-mass height $\sim$0.1~kpc. \rev{The disc height increases from $\sim$0.04~kpc in low-mass galaxies to $\sim$0.2~kpc in higher mass galaxies.} The disc height gradually increases with the age of stellar population, from $\sim$0.1~kpc to $\sim$0.8~kpc   (\autoref{fig:sum_zhalf}). There may be a weak trend for the disc height of young stars to increase with galaxy mass, but no systematic trend for stars older than 100~Myr (\autoref{fig:sumzhalf_Mstar}).

  \item The short to long axis ratio $z_{1/2}/x_{1/2}$ also increases with the age of stellar population, from $\sim$0.15 to $\sim$0.6 (\autoref{fig:sum_z2x}). There is a slight trend for the axis ratio of the oldest stars to decrease with galaxy mass (\autoref{fig:sumz2x_Mstar}). 

  \item We trace the same group of stars over consecutive simulation snapshots and calculate the evolution of their intrinsic axis ratio (\autoref{fig:sumtrace}). We confirm that stars form in thin discs $z_{1/2}/x_{1/2}\sim0.15$ and then rapidly expand away from the current disc plane.
  
  \item However, in addition to the real kinematic heating of stars, we find a new effect contributing to thicker appearance of galactic disks. The disc plane in observations of high-redshift galaxies is defined by rest-frame UV light dominated by young stars. The orientation of this plane rapidly varies by $\sim$20\degree-40\degree\ (\autoref{fig:anglechange}), which mixes and artificially inflates the configuration of older stars. The plane continues to precess after $\sim$200~Myr.
 \end{itemize}

In future studies it will be desirable to quantify the effect of disc precession using other simulations and at lower redshift. Since the mixing of stellar populations is rapid, with the timescale less than 200~Myr, this will require correspondingly high temporal resolution of the analyzed stellar populations. Previous studies advocating the "upside-down" disc formation scenario lacked sufficient time resolution to detect the plane precession effect. We plan to address these issues with upcoming galaxy formation simulations.

\section*{Acknowledgements}

We thank Hui Li for producing the simulation suite used in this study, and the referee for a very helpful report.
This work was supported in part by the US National Science Foundation through grant 1909063.

\section*{Data availability}
The data underlying this article will be shared on reasonable request to the corresponding author. 

\bibliographystyle{mnras}
\bibliography{bzd,gc,diskplus}

\begin{thebibliography}{}
\makeatletter
\relax
\def\mn@urlcharsother{\let\do\@makeother \do\$\do\&\do\#\do\^\do\_\do\%\do\~}
\def\mn@doi{\begingroup\mn@urlcharsother \@ifnextchar [ {\mn@doi@}
  {\mn@doi@[]}}
\def\mn@doi@[#1]#2{\def\@tempa{#1}\ifx\@tempa\@empty \href
  {http://dx.doi.org/#2} {doi:#2}\else \href {http://dx.doi.org/#2} {#1}\fi
  \endgroup}
\def\mn@eprint#1#2{\mn@eprint@#1:#2::\@nil}
\def\mn@eprint@arXiv#1{\href {http://arxiv.org/abs/#1} {{\tt arXiv:#1}}}
\def\mn@eprint@dblp#1{\href {http://dblp.uni-trier.de/rec/bibtex/#1.xml}
  {dblp:#1}}
\def\mn@eprint@#1:#2:#3:#4\@nil{\def\@tempa {#1}\def\@tempb {#2}\def\@tempc
  {#3}\ifx \@tempc \@empty \let \@tempc \@tempb \let \@tempb \@tempa \fi \ifx
  \@tempb \@empty \def\@tempb {arXiv}\fi \@ifundefined
  {mn@eprint@\@tempb}{\@tempb:\@tempc}{\expandafter \expandafter \csname
  mn@eprint@\@tempb\endcsname \expandafter{\@tempc}}}

\bibitem[\protect\citeauthoryear{{Bensby}, {Feltzing}, {Lundstr{\"o}m}  \&
  {Ilyin}}{{Bensby} et~al.}{2005}]{Bensby:2005aa}
{Bensby} T.,  {Feltzing} S.,  {Lundstr{\"o}m} I.,   {Ilyin} I.,  2005, \mn@doi
  [\aap] {10.1051/0004-6361:20040332}, \href
  {https://ui.adsabs.harvard.edu/abs/2005A&A...433..185B} {433, 185}

\bibitem[\protect\citeauthoryear{{Bensby}, {Alves-Brito}, {Oey}, {Yong}  \&
  {Mel{\'e}ndez}}{{Bensby} et~al.}{2011}]{Bensby:2011aa}
{Bensby} T.,  {Alves-Brito} A.,  {Oey} M.~S.,  {Yong} D.,   {Mel{\'e}ndez} J.,
  2011, \mn@doi [\apjl] {10.1088/2041-8205/735/2/L46}, \href
  {https://ui.adsabs.harvard.edu/abs/2011ApJ...735L..46B} {735, L46}

\bibitem[\protect\citeauthoryear{{Beraldo e Silva}, {Debattista},
  {Khachaturyants}  \& {Nidever}}{{Beraldo e Silva}
  et~al.}{2020}]{Beraldo-e-Silva:2020aa}
{Beraldo e Silva} L.,  {Debattista} V.~P.,  {Khachaturyants} T.,   {Nidever}
  D.,  2020, \mn@doi [\mnras] {10.1093/mnras/staa065}, \href
  {https://ui.adsabs.harvard.edu/abs/2020MNRAS.492.4716B} {492, 4716}

\bibitem[\protect\citeauthoryear{{Bird}, {Kazantzidis}, {Weinberg}, {Guedes},
  {Callegari}, {Mayer}  \& {Madau}}{{Bird} et~al.}{2013}]{Bird:2013aa}
{Bird} J.~C.,  {Kazantzidis} S.,  {Weinberg} D.~H.,  {Guedes} J.,  {Callegari}
  S.,  {Mayer} L.,   {Madau} P.,  2013, \mn@doi [\apj]
  {10.1088/0004-637X/773/1/43}, \href
  {https://ui.adsabs.harvard.edu/abs/2013ApJ...773...43B} {773, 43}

\bibitem[\protect\citeauthoryear{{Bird}, {Loebman}, {Weinberg}, {Brooks},
  {Quinn}  \& {Christensen}}{{Bird} et~al.}{2020}]{Bird:2020aa}
{Bird} J.~C.,  {Loebman} S.~R.,  {Weinberg} D.~H.,  {Brooks} A.,  {Quinn}
  T.~R.,   {Christensen} C.~R.,  2020, arXiv e-prints, \href
  {https://ui.adsabs.harvard.edu/abs/2020arXiv200512948B} {p. arXiv:2005.12948}

\bibitem[\protect\citeauthoryear{{Bournaud}, {Elmegreen}  \&
  {Martig}}{{Bournaud} et~al.}{2009}]{Bournaud:2009aa}
{Bournaud} F.,  {Elmegreen} B.~G.,   {Martig} M.,  2009, \mn@doi [\apjl]
  {10.1088/0004-637X/707/1/L1}, \href
  {https://ui.adsabs.harvard.edu/abs/2009ApJ...707L...1B} {707, L1}

\bibitem[\protect\citeauthoryear{{Bovy} \& {Rix}}{{Bovy} \&
  {Rix}}{2013}]{Bovy:2013aa}
{Bovy} J.,  {Rix} H.-W.,  2013, \mn@doi [\apj] {10.1088/0004-637X/779/2/115},
  \href {https://ui.adsabs.harvard.edu/abs/2013ApJ...779..115B} {779, 115}

\bibitem[\protect\citeauthoryear{{Bovy}, {Rix}, {Liu}, {Hogg}, {Beers}  \&
  {Lee}}{{Bovy} et~al.}{2012}]{Bovy:2012aa}
{Bovy} J.,  {Rix} H.-W.,  {Liu} C.,  {Hogg} D.~W.,  {Beers} T.~C.,   {Lee}
  Y.~S.,  2012, \mn@doi [\apj] {10.1088/0004-637X/753/2/148}, \href
  {https://ui.adsabs.harvard.edu/abs/2012ApJ...753..148B} {753, 148}

\bibitem[\protect\citeauthoryear{{Bovy}, {Rix}, {Schlafly}, {Nidever},
  {Holtzman}, {Shetrone}  \& {Beers}}{{Bovy} et~al.}{2016}]{Bovy:2016aa}
{Bovy} J.,  {Rix} H.-W.,  {Schlafly} E.~F.,  {Nidever} D.~L.,  {Holtzman}
  J.~A.,  {Shetrone} M.,   {Beers} T.~C.,  2016, \mn@doi [\apj]
  {10.3847/0004-637X/823/1/30}, \href
  {https://ui.adsabs.harvard.edu/abs/2016ApJ...823...30B} {823, 30}

\bibitem[\protect\citeauthoryear{{Buck}, {Obreja}, {Macci{\`o}}, {Minchev},
  {Dutton}  \& {Ostriker}}{{Buck} et~al.}{2020}]{Buck:2020ab}
{Buck} T.,  {Obreja} A.,  {Macci{\`o}} A.~V.,  {Minchev} I.,  {Dutton} A.~A.,
  {Ostriker} J.~P.,  2020, \mn@doi [\mnras] {10.1093/mnras/stz3241}, \href
  {https://ui.adsabs.harvard.edu/abs/2020MNRAS.491.3461B} {491, 3461}

\bibitem[\protect\citeauthoryear{{Ceverino}, {Primack}  \& {Dekel}}{{Ceverino}
  et~al.}{2015}]{Ceverino:2015aa}
{Ceverino} D.,  {Primack} J.,   {Dekel} A.,  2015, \mn@doi [\mnras]
  {10.1093/mnras/stv1603}, \href
  {http://adsabs.harvard.edu/abs/2015MNRAS.453..408C} {453, 408}

\bibitem[\protect\citeauthoryear{{Comer{\'o}n} et~al.,}{{Comer{\'o}n}
  et~al.}{2011}]{Comeron:2011aa}
{Comer{\'o}n} S.,  et~al., 2011, \mn@doi [\apj] {10.1088/0004-637X/741/1/28},
  \href {https://ui.adsabs.harvard.edu/abs/2011ApJ...741...28C} {741, 28}

\bibitem[\protect\citeauthoryear{{Comer{\'o}n}, {Elmegreen}, {Salo},
  {Laurikainen}, {Holwerda}  \& {Knapen}}{{Comer{\'o}n}
  et~al.}{2014}]{Comeron:2014aa}
{Comer{\'o}n} S.,  {Elmegreen} B.~G.,  {Salo} H.,  {Laurikainen} E.,
  {Holwerda} B.~W.,   {Knapen} J.~H.,  2014, \mn@doi [\aap]
  {10.1051/0004-6361/201424412}, \href
  {https://ui.adsabs.harvard.edu/abs/2014A&A...571A..58C} {571, A58}

\bibitem[\protect\citeauthoryear{{Elmegreen} \& {Elmegreen}}{{Elmegreen} \&
  {Elmegreen}}{2006}]{Elmegreen:2006aa}
{Elmegreen} B.~G.,  {Elmegreen} D.~M.,  2006, \mn@doi [\apj] {10.1086/507578},
  \href {https://ui.adsabs.harvard.edu/abs/2006ApJ...650..644E} {650, 644}

\bibitem[\protect\citeauthoryear{{Elmegreen}, {Elmegreen}, {Rubin}  \&
  {Schaffer}}{{Elmegreen} et~al.}{2005}]{Elmegreen:2005aa}
{Elmegreen} D.~M.,  {Elmegreen} B.~G.,  {Rubin} D.~S.,   {Schaffer} M.~A.,
  2005, \mn@doi [\apj] {10.1086/432502}, \href
  {http://adsabs.harvard.edu/abs/2005ApJ...631...85E} {631, 85}

\bibitem[\protect\citeauthoryear{{Elmegreen}, {Elmegreen}, {Ravindranath}  \&
  {Coe}}{{Elmegreen} et~al.}{2007}]{Elmegreen:2007aa}
{Elmegreen} D.~M.,  {Elmegreen} B.~G.,  {Ravindranath} S.,   {Coe} D.~A.,
  2007, \mn@doi [\apj] {10.1086/511667}, \href
  {http://adsabs.harvard.edu/abs/2007ApJ...658..763E} {658, 763}

\bibitem[\protect\citeauthoryear{{Elmegreen}, {Elmegreen}, {Tompkins}  \&
  {Jenks}}{{Elmegreen} et~al.}{2017}]{Elmegreen:2017aa}
{Elmegreen} B.~G.,  {Elmegreen} D.~M.,  {Tompkins} B.,   {Jenks} L.~G.,  2017,
  \mn@doi [\apj] {10.3847/1538-4357/aa88d4}, \href
  {http://adsabs.harvard.edu/abs/2017ApJ...847...14E} {847, 14}

\bibitem[\protect\citeauthoryear{{Gnedin}}{{Gnedin}}{2014}]{Gnedin:2014aa}
{Gnedin} N.~Y.,  2014, \mn@doi [\apj] {10.1088/0004-637X/793/1/29}, \href
  {http://adsabs.harvard.edu/abs/2014ApJ...793...29G} {793, 29}

\bibitem[\protect\citeauthoryear{{Gnedin} \& {Abel}}{{Gnedin} \&
  {Abel}}{2001}]{Gnedin:2001aa}
{Gnedin} N.~Y.,  {Abel} T.,  2001, \mn@doi [NewA]
  {10.1016/S1384-1076(01)00068-9}, \href
  {http://adsabs.harvard.edu/abs/2001NewA....6..437G} {6, 437}

\bibitem[\protect\citeauthoryear{{Gnedin} \& {Kravtsov}}{{Gnedin} \&
  {Kravtsov}}{2011}]{Gnedin:2011aa}
{Gnedin} N.~Y.,  {Kravtsov} A.~V.,  2011, \mn@doi [\apj]
  {10.1088/0004-637X/728/2/88}, \href
  {http://adsabs.harvard.edu/abs/2011ApJ...728...88G} {728, 88}

\bibitem[\protect\citeauthoryear{{Grand}, {Springel}, {G{\'o}mez}, {Marinacci},
  {Pakmor}, {Campbell}  \& {Jenkins}}{{Grand} et~al.}{2016}]{Grand:2016aa}
{Grand} R. J.~J.,  {Springel} V.,  {G{\'o}mez} F.~A.,  {Marinacci} F.,
  {Pakmor} R.,  {Campbell} D. J.~R.,   {Jenkins} A.,  2016, \mn@doi [\mnras]
  {10.1093/mnras/stw601}, \href
  {https://ui.adsabs.harvard.edu/abs/2016MNRAS.459..199G} {459, 199}

\bibitem[\protect\citeauthoryear{{Haardt} \& {Madau}}{{Haardt} \&
  {Madau}}{2001}]{Haardt:2001aa}
{Haardt} F.,  {Madau} P.,  2001, Clusters of Galaxies and the High Redshift
  Universe Observed in X-rays, \href
  {http://adsabs.harvard.edu/abs/2001cghr.confE..64H} {p.~64}

\bibitem[\protect\citeauthoryear{{Juri{\'c}} et~al.,}{{Juri{\'c}}
  et~al.}{2008}]{Juric:2008aa}
{Juri{\'c}} M.,  et~al., 2008, \mn@doi [\apj] {10.1086/523619}, \href
  {https://ui.adsabs.harvard.edu/abs/2008ApJ...673..864J} {673, 864}

\bibitem[\protect\citeauthoryear{{Kravtsov}}{{Kravtsov}}{1999}]{Kravtsov:1999aa}
{Kravtsov} A.~V.,  1999, PhD thesis, NEW MEXICO STATE UNIVERSITY

\bibitem[\protect\citeauthoryear{{Kravtsov}}{{Kravtsov}}{2003}]{Kravtsov:2003aa}
{Kravtsov} A.~V.,  2003, \mn@doi [\apjl] {10.1086/376674}, \href
  {http://adsabs.harvard.edu/abs/2003ApJ...590L...1K} {590, L1}

\bibitem[\protect\citeauthoryear{{Kravtsov}, {Klypin}  \&
  {Khokhlov}}{{Kravtsov} et~al.}{1997}]{Kravtsov:1997aa}
{Kravtsov} A.~V.,  {Klypin} A.~A.,   {Khokhlov} A.~M.,  1997, \mn@doi [\apjs]
  {10.1086/313015}, \href {http://adsabs.harvard.edu/abs/1997ApJS..111...73K}
  {111, 73}

\bibitem[\protect\citeauthoryear{{Lacey}}{{Lacey}}{1984}]{Lacey:1984aa}
{Lacey} C.~G.,  1984, \mn@doi [\mnras] {10.1093/mnras/208.4.687}, \href
  {https://ui.adsabs.harvard.edu/abs/1984MNRAS.208..687L} {208, 687}

\bibitem[\protect\citeauthoryear{{Law}, {Steidel}, {Shapley}, {Nagy}, {Reddy}
  \& {Erb}}{{Law} et~al.}{2012}]{Law:2012aa}
{Law} D.~R.,  {Steidel} C.~C.,  {Shapley} A.~E.,  {Nagy} S.~R.,  {Reddy} N.~A.,
    {Erb} D.~K.,  2012, \mn@doi [\apj] {10.1088/0004-637X/745/1/85}, \href
  {https://ui.adsabs.harvard.edu/abs/2012ApJ...745...85L} {745, 85}

\bibitem[\protect\citeauthoryear{{Li} \& {Gnedin}}{{Li} \&
  {Gnedin}}{2019}]{li_gnedin19}
{Li} H.,  {Gnedin} O.~Y.,  2019, \mn@doi [\mnras] {10.1093/mnras/stz1114},
  \href {http://adsabs.harvard.edu/abs/2019MNRAS.486.4030L} {486, 4030}

\bibitem[\protect\citeauthoryear{{Li}, {Gnedin}, {Gnedin}, {Meng}, {Semenov}
  \& {Kravtsov}}{{Li} et~al.}{2017}]{Li:2017ab}
{Li} H.,  {Gnedin} O.~Y.,  {Gnedin} N.~Y.,  {Meng} X.,  {Semenov} V.~A.,
  {Kravtsov} A.~V.,  2017, \mn@doi [\apj] {10.3847/1538-4357/834/1/69}, \href
  {http://adsabs.harvard.edu/abs/2017ApJ...834...69L} {834, 69}

\bibitem[\protect\citeauthoryear{{Li}, {Gnedin}  \& {Gnedin}}{{Li}
  et~al.}{2018}]{Li:2018aa}
{Li} H.,  {Gnedin} O.~Y.,   {Gnedin} N.~Y.,  2018, \mn@doi [\apj]
  {10.3847/1538-4357/aac9b8}, \href
  {http://adsabs.harvard.edu/abs/2018ApJ...861..107L} {861, 107}

\bibitem[\protect\citeauthoryear{{Loebman}, {Ro{\v{s}}kar}, {Debattista},
  {Ivezi{\'c}}, {Quinn}  \& {Wadsley}}{{Loebman} et~al.}{2011}]{Loebman:2011aa}
{Loebman} S.~R.,  {Ro{\v{s}}kar} R.,  {Debattista} V.~P.,  {Ivezi{\'c}}
  {\v{Z}}.,  {Quinn} T.~R.,   {Wadsley} J.,  2011, \mn@doi [\apj]
  {10.1088/0004-637X/737/1/8}, \href
  {https://ui.adsabs.harvard.edu/abs/2011ApJ...737....8L} {737, 8}

\bibitem[\protect\citeauthoryear{{Ma}, {Hopkins}, {Wetzel}, {Kirby},
  {Angl{\'e}s-Alc{\'a}zar}, {Faucher-Gigu{\`e}re}, {Kere{\v{s}}}  \&
  {Quataert}}{{Ma} et~al.}{2017}]{Ma:2017aa}
{Ma} X.,  {Hopkins} P.~F.,  {Wetzel} A.~R.,  {Kirby} E.~N.,
  {Angl{\'e}s-Alc{\'a}zar} D.,  {Faucher-Gigu{\`e}re} C.-A.,  {Kere{\v{s}}} D.,
    {Quataert} E.,  2017, \mn@doi [\mnras] {10.1093/mnras/stx273}, \href
  {https://ui.adsabs.harvard.edu/abs/2017MNRAS.467.2430M} {467, 2430}

\bibitem[\protect\citeauthoryear{{Martizzi}, {Faucher-Gigu{\`e}re}  \&
  {Quataert}}{{Martizzi} et~al.}{2015}]{Martizzi:2015aa}
{Martizzi} D.,  {Faucher-Gigu{\`e}re} C.-A.,   {Quataert} E.,  2015, \mn@doi
  [\mnras] {10.1093/mnras/stv562}, \href
  {http://adsabs.harvard.edu/abs/2015MNRAS.450..504M} {450, 504}

\bibitem[\protect\citeauthoryear{{Meng} \& {Gnedin}}{{Meng} \&
  {Gnedin}}{2020}]{Meng:2020aa}
{Meng} X.,  {Gnedin} O.~Y.,  2020, \mn@doi [\mnras] {10.1093/mnras/staa776},
  \href {https://ui.adsabs.harvard.edu/abs/2020MNRAS.494.1263M} {494, 1263}

\bibitem[\protect\citeauthoryear{{Meng}, {Gnedin}  \& {Li}}{{Meng}
  et~al.}{2019}]{Meng:2019aa}
{Meng} X.,  {Gnedin} O.~Y.,   {Li} H.,  2019, \mn@doi [\mnras]
  {10.1093/mnras/stz925}, \href
  {https://ui.adsabs.harvard.edu/abs/2019MNRAS.486.1574M} {486, 1574}

\bibitem[\protect\citeauthoryear{{Navarro} et~al.,}{{Navarro}
  et~al.}{2018}]{Navarro:2018aa}
{Navarro} J.~F.,  et~al., 2018, \mn@doi [\mnras] {10.1093/mnras/sty497}, \href
  {https://ui.adsabs.harvard.edu/abs/2018MNRAS.476.3648N} {476, 3648}

\bibitem[\protect\citeauthoryear{{Overzier}, {Heckman}, {Schiminovich},
  {Basu-Zych}, {Gon{\c c}alves}, {Martin}  \& {Rich}}{{Overzier}
  et~al.}{2010}]{Overzier:2010aa}
{Overzier} R.~A.,  {Heckman} T.~M.,  {Schiminovich} D.,  {Basu-Zych} A.,
  {Gon{\c c}alves} T.,  {Martin} D.~C.,   {Rich} R.~M.,  2010, \mn@doi [\apj]
  {10.1088/0004-637X/710/2/979}, \href
  {http://adsabs.harvard.edu/abs/2010ApJ...710..979O} {710, 979}

\bibitem[\protect\citeauthoryear{{Pillepich} et~al.,}{{Pillepich}
  et~al.}{2019}]{Pillepich:2019aa}
{Pillepich} A.,  et~al., 2019, \mn@doi [\mnras] {10.1093/mnras/stz2338}, \href
  {https://ui.adsabs.harvard.edu/abs/2019MNRAS.490.3196P} {490, 3196}

\bibitem[\protect\citeauthoryear{{Reddy}, {Lambert}  \& {Allende
  Prieto}}{{Reddy} et~al.}{2006}]{Reddy:2006aa}
{Reddy} B.~E.,  {Lambert} D.~L.,   {Allende Prieto} C.,  2006, \mn@doi [\mnras]
  {10.1111/j.1365-2966.2006.10148.x}, \href
  {https://ui.adsabs.harvard.edu/abs/2006MNRAS.367.1329R} {367, 1329}

\bibitem[\protect\citeauthoryear{{Reshetnikov}, {Dettmar}  \&
  {Combes}}{{Reshetnikov} et~al.}{2003}]{Reshetnikov:2003aa}
{Reshetnikov} V.~P.,  {Dettmar} R.~J.,   {Combes} F.,  2003, \mn@doi [\aap]
  {10.1051/0004-6361:20021874}, \href
  {https://ui.adsabs.harvard.edu/abs/2003A&A...399..879R} {399, 879}

\bibitem[\protect\citeauthoryear{{Rudd}, {Zentner}  \& {Kravtsov}}{{Rudd}
  et~al.}{2008}]{Rudd:2008aa}
{Rudd} D.~H.,  {Zentner} A.~R.,   {Kravtsov} A.~V.,  2008, \mn@doi [\apj]
  {10.1086/523836}, \href {http://adsabs.harvard.edu/abs/2008ApJ...672...19R}
  {672, 19}

\bibitem[\protect\citeauthoryear{{Schmidt} et~al.,}{{Schmidt}
  et~al.}{2014}]{Schmidt:2014aa}
{Schmidt} W.,  et~al., 2014, \mn@doi [\mnras] {10.1093/mnras/stu501}, \href
  {http://adsabs.harvard.edu/abs/2014MNRAS.440.3051S} {440, 3051}

\bibitem[\protect\citeauthoryear{{Sch{\"o}nrich} \& {Binney}}{{Sch{\"o}nrich}
  \& {Binney}}{2009}]{Schonrich:2009aa}
{Sch{\"o}nrich} R.,  {Binney} J.,  2009, \mn@doi [\mnras]
  {10.1111/j.1365-2966.2009.15365.x}, \href
  {https://ui.adsabs.harvard.edu/abs/2009MNRAS.399.1145S} {399, 1145}

\bibitem[\protect\citeauthoryear{{Semenov}, {Kravtsov}  \& {Gnedin}}{{Semenov}
  et~al.}{2016}]{Semenov:2016aa}
{Semenov} V.~A.,  {Kravtsov} A.~V.,   {Gnedin} N.~Y.,  2016, \mn@doi [\apj]
  {10.3847/0004-637X/826/2/200}, \href
  {http://adsabs.harvard.edu/abs/2016ApJ...826..200S} {826, 200}

\bibitem[\protect\citeauthoryear{{Spitzer} \& {Schwarzschild}}{{Spitzer} \&
  {Schwarzschild}}{1951}]{Spitzer:1951aa}
{Spitzer} Lyman J.,  {Schwarzschild} M.,  1951, \mn@doi [\apj]
  {10.1086/145478}, \href
  {https://ui.adsabs.harvard.edu/abs/1951ApJ...114..385S} {114, 385}

\bibitem[\protect\citeauthoryear{{Spitzer} \& {Schwarzschild}}{{Spitzer} \&
  {Schwarzschild}}{1953}]{Spitzer:1953aa}
{Spitzer} Lyman J.,  {Schwarzschild} M.,  1953, \mn@doi [\apj]
  {10.1086/145730}, \href
  {https://ui.adsabs.harvard.edu/abs/1953ApJ...118..106S} {118, 106}

\bibitem[\protect\citeauthoryear{{Stinson} et~al.,}{{Stinson}
  et~al.}{2013}]{Stinson:2013aa}
{Stinson} G.~S.,  et~al., 2013, \mn@doi [\mnras] {10.1093/mnras/stt1600}, \href
  {https://ui.adsabs.harvard.edu/abs/2013MNRAS.436..625S} {436, 625}

\bibitem[\protect\citeauthoryear{{Swinbank} et~al.,}{{Swinbank}
  et~al.}{2010}]{Swinbank:2010aa}
{Swinbank} A.~M.,  et~al., 2010, \mn@doi [\mnras]
  {10.1111/j.1365-2966.2010.16485.x}, \href
  {http://adsabs.harvard.edu/abs/2010MNRAS.405..234S} {405, 234}

\bibitem[\protect\citeauthoryear{{Tutukov}, {Shustov}  \& {Wiebe}}{{Tutukov}
  et~al.}{2000}]{Tutukov:2000aa}
{Tutukov} A.~V.,  {Shustov} B.~M.,   {Wiebe} D.~S.,  2000, \mn@doi [Astronomy
  Reports] {10.1134/1.1320496}, \href
  {https://ui.adsabs.harvard.edu/abs/2000ARep...44..711T} {44, 711}

\bibitem[\protect\citeauthoryear{{Yoachim} \& {Dalcanton}}{{Yoachim} \&
  {Dalcanton}}{2008a}]{Yoachim:2008aa}
{Yoachim} P.,  {Dalcanton} J.~J.,  2008a, \mn@doi [\apj] {10.1086/589553},
  \href {https://ui.adsabs.harvard.edu/abs/2008ApJ...682.1004Y} {682, 1004}

\bibitem[\protect\citeauthoryear{{Yoachim} \& {Dalcanton}}{{Yoachim} \&
  {Dalcanton}}{2008b}]{Yoachim:2008ab}
{Yoachim} P.,  {Dalcanton} J.~J.,  2008b, \mn@doi [\apj] {10.1086/590246},
  \href {https://ui.adsabs.harvard.edu/abs/2008ApJ...683..707Y} {683, 707}

\bibitem[\protect\citeauthoryear{{Zemp}, {Gnedin}, {Gnedin}  \&
  {Kravtsov}}{{Zemp} et~al.}{2011}]{zemp_etal11}
{Zemp} M.,  {Gnedin} O.~Y.,  {Gnedin} N.~Y.,   {Kravtsov} A.~V.,  2011, \mn@doi
  [\apjs] {10.1088/0067-0049/197/2/30}, \href
  {http://adsabs.harvard.edu/abs/2011ApJS..197...30Z} {197, 30}

\bibitem[\protect\citeauthoryear{{Zhang} et~al.,}{{Zhang}
  et~al.}{2019}]{Zhang:2019aa}
{Zhang} H.,  et~al., 2019, \mn@doi [\mnras] {10.1093/mnras/stz339}, \href
  {http://adsabs.harvard.edu/abs/2019MNRAS.484.5170Z} {484, 5170}

\bibitem[\protect\citeauthoryear{{van der Wel} et~al.,}{{van der Wel}
  et~al.}{2014}]{van-der-Wel:2014aa}
{van der Wel} A.,  et~al., 2014, \mn@doi [\apjl] {10.1088/2041-8205/792/1/L6},
  \href {http://adsabs.harvard.edu/abs/2014ApJ...792L...6V} {792, L6}

\makeatother
\end{thebibliography}

% Don't change these lines
\bsp	% typesetting comment
\label{lastpage}
\end{document}